\documentclass[twocolumn,preprintnumbers,pra,hyperref]{revtex4}
\usepackage{amssymb}
\usepackage{amsfonts}
\usepackage{amsmath}
\usepackage{graphicx}

\begin{document}

\title{Entanglement and nonclassicality of photon-added two-mode squeezed
thermal state}
\author{Li-Yun Hu$^{1,2,\dag }$\thanks{{\small E-mail: hlyun2008@126.com.}},
Fang Jia$^{1}$ and Zhi-Ming Zhang$^{2,\ast }$\thanks{{\small E-mail:
zmzhang@scnu.edu.cn}}}
\affiliation{$^{1}${\small Department of physics, Jiangxi Normal University, Nanchang
330022, China}\\
$^{2}${\small Laboratory of Nanophotonic Functional Materials and Devices,
SIPSE \& LQIT, South China Normal University, Guangzhou 510006, China}\\
$\dag ${\small Email: hlyun@jsnu.edu.cn; }$\ast $ {\small Email:
zmzhang@scnu.edu.cn.}}

\begin{abstract}
{\small We introduce a kind of entangled state---photon-addition two-mode
squeezed thermal state (TMSTS) by adding photons to each mode of the TMSTS.
Using the P-representation of thermal state, the compact expression of the
normalization factor is derived, a Jacobi polynomial.} {\small The
nonclassicality is investigated by exploring especially the negativity of
Wigner function. The entanglement is discussed by using Shchukin-Vogel
criteria. It is shown that the photon-addtion to the TMSTS may be more
effective for the entanglement enhancement than the photon-subtraction from
the TMSTS. In addition, the quantum teleportation is also examined, which
shows that symmetrical photon-added TMSTS may be more useful for quantum
teleportation than the non-symmetric case.}

PACS number(s): 42.50.Dv, 03.65.Wj, 03.67.Mn
\end{abstract}

\maketitle

\section{Introduction}

Quantum entanglement with continuous-variable is an essential resource in
quantum information processing \cite{1}, such as teleportation, dense
coding, and quantum cloning. In a quantum optics laboratory, a Gaussian
two-mode squeezed vacuum state is ofen used as entangled resource, which
cannot be distilled only by Gaussian local operators and classical
communications due to the limitation from the no-go theorem \cite{2,3,4}. To
satisfy the requirement of quantum information protocols for long-distance
communication, there have been suggestions and realizations for engineering
the quantum state, which are plausible ways to conditionally manipulate a
nonclassical state of an optical field by subtracting or adding photons
from/to a Gaussian field \cite{5,6,7,8,8a,9,10}. Actually, the photon
addition and subtraction have been successfully demonstrated experimentally
for probing quantum commutation rules by Parigi et al. \cite{11}.

In order to increase quantum entanglement, two-mode photon-subtraction
squeezed vacuum states (TPSSV) have received more attention from both
experimentalists and theoreticians \cite{5,9,12,13,14,15,16,17,18,19,20,21}.
Olivares et al. \cite{12} considered the photon subtraction using on--off
photo detectors and showed improvement of quantum teleportation, depending
on the various parameters involved. Kitagawa et al. \cite{13}, on the other
hand, investigated the degree of entanglement for the TPSSV by using an
on--off photon detector. Using an operation with single photon counts,
Ourjoumtsev et al. \cite{14,15} demonstrated experimentally that
entanglement between Gaussian entangled states can be increased by
subtracting only one photon from two-mode squeezed vacuum states. In
addition, Lee et. al \cite{21} proposed a coherent superposition of photon
subtraction and addition to enhance quantum entanglement of two-mode
Gaussian sate. It is shown that, especially for the small-squeezing regime,
the effects of coherent operation are more prominent than those of the mere
photon subtraction and the photon addition.

Recently, we proposed the any photon-added squeezed thermal state
theoretically, and investigated its nonclassicality by exploring the
sub-Poissonian and negative Wigner function (WF) \cite{22}. The results show
that the WF of single photon-added squeezed thermal state (PASTS) always has
negative values at the phase space center. The decoherence effect on the
PASTS is examined by the analytical expression of WF. It is found that a
longer threshold value of decay time is included in single PASTS than in
single-photon subtraction squeezed thermal state (STS). In this paper, as a
natural extension, we shall introduce a kind of nonclassical
state---photon-addition two-mode STS (PA-TMSTS), generated by adding photons
to each mode of two-mode STS (TMSTS) which can be considered as a
generalized bipartite Gaussian state. Then we shall investigate the
entanglement and nonclassical properties.

This paper is organized as follows. In Sec. II we introduce the PA-TMSTS. By
using the P-representation of density operator of thermal state, we derive
the normal ordering and anti-normal form of the TMSTS, which is convenient to
obtain distribution function, such as Q-function and WF. Then a compact
expression for the normalization factor of the PA-TMSTS, which is a Jacobi
polynomial of squeezing parameter $r$ and mean number $\bar{n}$ of thermal
state. In Sec III, we present the nonclassical properties of the PA-TMSTS in
terms of cross-correlation function, distribution of photon number,
antibunching effect and the negativity of its WF. It is shown that the WF
lost its Gaussian property in phase space due to the presence of
two-variable Hermite polynomials and the WF of single PA-TMSTS always has
its negative region at the center of phase space. Then, in Secs. IV and V
are devoted to discussing the entanglement properties of the PA-TMSTS by
Shchukin-Vogel criteria and the quantum teleportation. The conclusions are
involved in Sec. VI.

\section{Photon-addition two-mode squeezed thermal state (PA-TMSTS)}

As Agarwal et al \cite{23}. introduced the excitations on a coherent state
by repeated application of the photon creation operator on the coherent
state, we introduce theoretically the photon-addition two-mode squeezed
thermal state (PA-TMSTS).

For two-mode case, the photon-added scheme can be presented by the mapping $%
\rho \rightarrow a^{\dag m}b^{\dag n}\rho a^{m}b^{\dag n}$. Here we
introduce the PA-TMSTS, which can be generated by repeatedly operating the
photon creation operator $a^{\dagger }$ and $b^{\dagger }$ on a two-mode
squeezed thermal state (TMSTS), so its density operator is%
\begin{equation}
\rho ^{SA}\equiv N_{m,n}^{-1}{}a^{\dagger m}b^{\dagger n}S\left( r\right)
\rho _{th1}\rho _{th2}S^{\dagger }\left( r\right) a^{m}b^{n},  \label{t1}
\end{equation}%
where $m,n$ are the added photon number to each mode (non-negative
integers), and $N_{m,n}$ is the normalization of the PA-TMSTS to be
determined by $\mathtt{tr}\rho ^{SA}=1$, and $S(r)=\exp [r(a^{\dagger
}b^{\dagger }-ab)]$ is the two-mode squeezing operator with squeezing
parameter $r$. Here $\rho _{th1,2}$ is a density operator of single-mode
thermal state,%
\begin{equation}
\rho _{th1,2}=\sum_{n=0}^{\infty }\frac{\bar{n}^{n}}{\left( \bar{n}+1\right)
^{n+1}}\left\vert n\right\rangle \left\langle n\right\vert ,  \label{t2}
\end{equation}%
where $\bar{n}$ is the average photon number of thermal state $\rho _{thj}$ (%
$j=1,2$). For simplicity, we assume the average photon number of $\rho
_{thj} $ ($j=1,2$) to be identical. In addition, the P-representation of
density operator $\rho _{thj}$ can be expanded as \cite{24}%
\begin{equation}
\rho _{thj}=\frac{1}{\bar{n}}\int \frac{d^{2}\alpha }{\pi }e^{-\frac{1}{\bar{%
n}}\left\vert \alpha \right\vert ^{2}}\left\vert \alpha \right\rangle
\left\langle \alpha \right\vert ,  \label{t3}
\end{equation}%
which is useful for later calculation and here $\left\vert \alpha
\right\rangle $ is the coherent state.

\subsection{Normal ordering and anti-normal form of the TMSTS}

In order to simplify our calculation, here we shall derive the normally
ordering form of the TMSTS. For this purpose, we examine the two-mode
squeezed coherent states $S\left\vert \alpha ,\beta \right\rangle $ ($%
\left\vert \alpha ,\beta \right\rangle =\left\vert \alpha \right\rangle
\otimes \left\vert \beta \right\rangle $). Note that $\left\vert \alpha
\right\rangle =\exp [-\frac{1}{2}\left\vert \alpha \right\vert ^{2}+\alpha
a^{\dagger }]\left\vert 0\right\rangle $ and the following transformation
relations \cite{25,26}:
\begin{align}
S(r)a^{\dagger }S^{\dagger }(r)& =a^{\dagger }\cosh r-b\sinh r,  \notag \\
S(r)b^{\dagger }S^{\dagger }(r)& =b^{\dagger }\cosh r-a\sinh r,  \label{t4}
\end{align}%
we see%
\begin{eqnarray}
S\left\vert \alpha ,\beta \right\rangle &=&\text{sech}r\exp \left[ -\frac{1}{%
2}(\left\vert \alpha \right\vert ^{2}+\left\vert \beta \right\vert ^{2})%
\right]  \notag \\
&&\times \exp \left[ \alpha \left( a^{\dagger }\cosh r-b\sinh r\right) %
\right]  \notag \\
&&\times \exp [\beta \left( b^{\dagger }\cosh r-a\sinh r\right) ]  \notag \\
&&\times \exp \left[ a^{\dagger }b^{\dagger }\tanh r\right] \left\vert
00\right\rangle ,  \label{t5}
\end{eqnarray}%
where $S(r)\left\vert 00\right\rangle =$sech$r\exp \left[ a^{\dagger
}b^{\dagger }\tanh r\right] \left\vert 00\right\rangle $\ is used.

Further noting $e^{\tau a}a^{\dagger }e^{-\tau a}=a^{\dagger }+\tau ,$ and
for operators $A,B$ satisfying the conditions $\left[ A,[A,B]\right] =\left[
B,[A,B]\right] =0,$ we have $%
e^{A+B}=e^{A}e^{B}e^{-[A,B]/2}=e^{B}e^{A}e^{[A,B]/2},$ thus Eq.(\ref{t5})
can be put into the following form%
\begin{eqnarray}
S\left\vert \alpha ,\beta \right\rangle &=&\text{sech}r\exp \left[ -\frac{1}{%
2}(\left\vert \alpha \right\vert ^{2}+\left\vert \beta \right\vert
^{2})-\alpha \beta \tanh r\right]  \notag \\
&&\times \exp \left[ \left( a^{\dagger }\alpha +b^{\dagger }\beta \right)
\text{sech}r+a^{\dagger }b^{\dagger }\tanh r\right] \left\vert
00\right\rangle .  \label{t6}
\end{eqnarray}%
Thus inserting Eq.(\ref{t6}) into Eq.(\ref{t3}) and using the vacuum
projector $\left\vert 00\right\rangle \left\langle 00\right\vert =\colon
\exp [-a^{\dagger }a-b^{\dagger }b]\colon $ (where $\colon \colon $ denotes
the normally ordering) as well as the IWOP technique \cite{27,28}, we can
obtain%
\begin{eqnarray}
\rho ^{S} &\equiv &S\rho _{th1}\rho _{th2}S^{\dagger }  \notag \\
&=&\frac{1}{\bar{n}^{2}}\int \frac{d^{2}\alpha d^{2}\beta }{\pi ^{2}}e^{-%
\frac{1}{\bar{n}}(\left\vert \alpha \right\vert ^{2}+\left\vert \beta
\right\vert ^{2})}S\left\vert \alpha ,\beta \right\rangle \left\langle
\alpha ,\beta \right\vert S^{\dagger }  \notag \\
&=&A_{1}\colon \exp \left[ A_{2}\left( a^{\dagger }b^{\dagger }+ab\right)
-A_{3}\left( a^{\dagger }a+b^{\dagger }b\right) \right] \colon ,  \label{t7}
\end{eqnarray}%
where we have set
\begin{eqnarray}
A_{1} &=&\frac{\text{sech}^{2}r}{\left( \bar{n}+1\right) ^{2}-\bar{n}%
^{2}\tanh ^{2}r},  \notag \\
A_{2} &=&\frac{\left( 2\bar{n}+1\right) \sinh r\cosh r}{\left( 2\bar{n}%
+\allowbreak 1\right) \cosh ^{2}r+\bar{n}^{2}},  \notag \\
A_{3} &=&\frac{\allowbreak \cosh ^{2}r+\bar{n}\cosh 2r}{\left( 2\bar{n}%
+\allowbreak 1\right) \cosh ^{2}r+\bar{n}^{2}},  \label{t8}
\end{eqnarray}%
and used the integration formula \cite{29}%
\begin{equation}
\int \frac{d^{2}z}{\pi }e^{\zeta \left\vert z\right\vert ^{2}+\xi z+\eta
z^{\ast }}=-\frac{1}{\zeta }e^{-\frac{\xi \eta }{\zeta }},\text{Re}\zeta <0.
\label{t9}
\end{equation}%
Eq.(\ref{t7}) is just the normally ordering form of TMSTS to be used to
realize our calculations below.

In addition, using Eqs.(\ref{t7}), (\ref{t9}) and the formula converting any
single-mode operator $\hat{O}$ into its anti-normal ordering form \cite{30},%
\begin{equation}
\hat{O}=\vdots \int \frac{d^{2}z}{\pi }\left\langle -z\right\vert \hat{O}%
\left\vert z\right\rangle e^{|z|^{2}+z^{\ast }a-za^{\dagger }+a^{\dagger
}a}\vdots ,  \label{t10}
\end{equation}%
where $\left\vert z\right\rangle $ is the coherent state, and the symbol $%
\vdots $ $\vdots $ denotes antinormal ordering, (note that the order of Bose
operators $a$ and $a^{\dagger }$ within $\vdots $ $\vdots $ can be
permuted), one can obtain the anti-normal ordering form of the TMSTS,%
\begin{equation}
\rho ^{S}=\tilde{A}_{1}\vdots \exp \left[ \tilde{A}_{2}\left( a^{\dagger
}b^{\dagger }+ab\right) -\tilde{A}_{3}\left( a^{\dagger }a+b^{\dagger
}b\right) \right] \vdots ,  \label{t11}
\end{equation}%
where we have set
\begin{eqnarray}
\tilde{A}_{1} &=&\frac{1}{\left( \bar{n}+1\right) ^{2}-\left( 2\bar{n}%
+1\right) \cosh ^{2}r},  \notag \\
\tilde{A}_{2} &=&\frac{\left( 2n+1\right) \sinh r\cosh r}{\left( \bar{n}%
+1\right) ^{2}-\left( 2\bar{n}+1\right) \cosh ^{2}r},  \notag \\
\tilde{A}_{3} &=&\frac{\sinh ^{2}r+n\cosh 2r}{\left( \bar{n}+1\right)
^{2}-\left( 2\bar{n}+1\right) \cosh ^{2}r}.  \label{t12}
\end{eqnarray}%
Eq.(\ref{t11}) implies that the P function $P(\alpha ,\beta )$ of the TMSTS
is
\begin{equation}
P(\alpha ,\beta )=\tilde{A}_{1}\exp \left[ \tilde{A}_{2}\left( \alpha ^{\ast
}\beta ^{\ast }+\alpha \beta \right) -\tilde{A}_{3}(\left\vert \alpha
\right\vert ^{2}+\left\vert \beta \right\vert ^{2})\right] ,  \label{t13}
\end{equation}%
which leads to the P representation of density operator $S\rho _{th1}\rho
_{th2}S^{\dagger }$ i.e.,%
\begin{equation}
\rho ^{S}=\int \frac{d^{2}\alpha d^{2}\beta }{\pi ^{2}}P(\alpha ,\beta
)\left\vert \alpha ,\beta \right\rangle \left\langle \alpha ,\beta
\right\vert .  \label{t14}
\end{equation}

In particular, for the case without squeezing, $r=0,$ then Eqs.(\ref{t7})
and (\ref{t11}) just reduce to, respectively,%
\begin{eqnarray}
\rho ^{S}\left( r=0\right) &=&\frac{1}{\left( \bar{n}+1\right) ^{2}}\colon
\exp \left[ -\frac{a^{\dagger }a+b^{\dagger }b}{\bar{n}+1}\right] \colon
\notag \\
&=&\frac{1}{\bar{n}^{2}}\vdots \exp \left[ -\frac{1}{\bar{n}}\left(
a^{\dagger }a+b^{\dagger }b\right) \right] \vdots ,  \label{t15}
\end{eqnarray}%
as expected \cite{24}. It is interesting to notice that, for the case of $%
\bar{n}=0$, corresponding to the two-mode squeezed vaccum state (TMSVS),
Eqs.(\ref{t7}) and (\ref{t11}) become
\begin{eqnarray}
&&\rho ^{S}\left( \bar{n}=0\right)  \notag \\
&=&\text{sech}^{2}r\colon \exp \left[ \left( a^{\dagger }b^{\dagger
}+ab\right) \tanh r-\left( a^{\dagger }a+b^{\dagger }b\right) \right] \colon
\notag \\
&=&-\text{csch}^{2}r\vdots \exp \left[ a^{\dagger }a+b^{\dagger }b-\left(
a^{\dagger }b^{\dagger }+ab\right) \coth r\right] \vdots ,  \label{t16}
\end{eqnarray}%
which are just the normal ordering form and anti-normal ordering form of the
TMSVS. The second equation in Eq.(\ref{t16}) seems a new result. Here, we
should mention that the normal (anti-)normal ordering forms of the TMSTS are
useful to higher-order squeezing and photon statistics \cite{31,32} for the
TMSTS.

\subsection{Normalization of the PA-TMSTS}

To fully describe a quantum state, its normalization is usually necessary.
Using Eq.(\ref{t7}), the PA-TMSTS reads as%
\begin{equation}
\rho ^{SA}=\frac{A_{1}}{N_{m,n}}\colon a^{\dagger m}b^{\dagger
n}e^{A_{2}\left( a^{\dagger }b^{\dagger }+ab\right) -A_{3}\left( a^{\dagger
}a+b^{\dagger }b\right) }a^{m}b^{n}\colon .  \label{t17}
\end{equation}%
Thus using the completeness relation of coherent state $\int d^{2}\alpha
d^{2}\beta \left\vert \alpha ,\beta \right\rangle \left\langle \alpha ,\beta
\right\vert /\pi ^{2}=1$ and Eq.(\ref{t9}), the normalization factor $%
N_{m,n} $ is given by (Appendix A)%
\begin{equation}
N_{m,n}=\left. \frac{\partial ^{2m+2n}}{\partial \tau ^{m}\partial
t^{m}\partial \tau ^{\prime n}\partial t^{\prime n}}e^{B_{1}\left( \tau
t+\tau ^{\prime }t^{\prime }\right) +B_{2}\left( \tau \tau ^{\prime
}+tt^{\prime }\right) }\right\vert _{t,\tau ,t^{\prime },\tau ^{\prime }=0},
\label{t18}
\end{equation}%
where
\begin{eqnarray}
B_{1} &=&\cosh ^{2}r+\bar{n}\cosh 2r,  \notag \\
B_{2} &=&\left( 2\bar{n}+1\right) \sinh r\cosh r.  \label{t19}
\end{eqnarray}%
Here we introduce a new expression of generating function for Jacobi
polynomials in form (Proof see Appendix B)%
\begin{align}
& \left. \frac{\partial ^{2m+2n}}{\partial \tau ^{m}\partial t^{m}\partial
\tau ^{\prime n}\partial t^{\prime n}}e^{A\left( \tau ^{\prime }t^{\prime
}+\tau t\right) +B\left( \tau \tau ^{\prime }+t^{\prime }t\right)
}\right\vert _{t,\tau ,t^{\prime },\tau ^{\prime }=0}  \notag \\
& =m!n!\left\{
\begin{array}{cc}
A^{n-m}\left( B^{2}-A^{2}\right) ^{m}P_{m}^{(n-m,0)}\left( \frac{B^{2}+A^{2}%
}{B^{2}-A^{2}}\right) & m\leqslant n \\
&  \\
A^{m-n}\left( B^{2}-A^{2}\right) ^{n}P_{n}^{(m-n,0)}\left( \frac{B^{2}+A^{2}%
}{B^{2}-A^{2}}\right) & n\leqslant m%
\end{array}%
\right. ,  \label{t20}
\end{align}%
thus the normalization factor $N_{m,n}$ can be put into (without loss of
generality assuming $m\leqslant n$)
\begin{equation}
N_{m,n}=m!n!B_{1}^{n-m}\omega ^{m}P_{m}^{(0,n-m)}\left( \frac{\upsilon }{%
\omega }\right) ,  \label{t21}
\end{equation}%
where we have used the property of the Jacobi polynomials $P_{m}^{(\alpha
,\beta )}(-x)=(-1)^{m}P_{m}^{(\beta ,\alpha )}(x),$and
\begin{eqnarray}
\omega &=&\bar{n}^{2}+\left( 2\bar{n}+1\right) \cosh ^{2}r,  \notag \\
\upsilon &=&\bar{n}\left( \bar{n}+1\right) \cosh 4r+\left( \allowbreak \bar{n%
}+\cosh ^{2}r\right) \cosh 2r.  \label{t22}
\end{eqnarray}%
Eq.(\ref{t21}) indicates that the normalization factor is related to the
Jacobi polynomials, which is important for further studying analytically the
statistical properties of the PA-TMSTS. Note Eq.(\ref{t21}) exhibits the
exchanging symmetry.

It is clear that, when $m=n=0,$ Eq.(\ref{t21}) just reduces to the TMSTS due
to $P_{0}^{(0,0)}\left( x\right) =1$; while for $n\neq 0$ and $m=0,$
noticing $P_{0}^{(0,n)}\left( x\right) =1,$ Eq.(\ref{t21}) becomes $%
N_{0,n}=n!B_{1}^{n}$. For the case $m=n$, $N_{m,m}$ is related to Legendre
polynomial of the parameter $\frac{\upsilon }{\omega }$, because of $%
P_{n}^{(0,0)}(x)=P_{n}(x),$ $P_{0}(x)=1$. In addition, when $\bar{n}=0$
leading to $\omega =B_{1}=\cosh ^{2}r\ $and $\frac{\upsilon }{\omega }=\cosh
2r,$ then Eq.(\ref{t21}) reads
\begin{equation}
N_{m,n}\left( \bar{n}=0\right) =m!n!\cosh ^{2n}rP_{m}^{(0,n-m)}\left( \cosh
2r\right) ,  \label{t23}
\end{equation}%
which is just the normalization of two-mode photon-added squeezed vacuum
state \cite{33}.

\section{Nonclassical properties of the PA-TMSTS}

In this section, we shall discuss the nonclassical properties of the
PA-TMSTS in terms of cross-correlation function, photon statistics,
anti-bunching effect and the negativity of its WF.

\subsection{Cross-correlation function of the PA-TMSTS}

The cross-correlation between the two modes reflects correlation between
photons in two different modes, which plays a key role in rendering many
two-mode radiations nonclassically. From Eqs. (\ref{t17}) and (\ref{t21}) we
can easily calculate the average photon number in the PA-TMSTS,%
\begin{equation}
\left\langle a^{\dagger }a\right\rangle =\frac{N_{m+1,n}}{N_{m,n}}%
-1,\left\langle b^{\dagger }b\right\rangle =\frac{N_{m,n+1}}{N_{m,n}}-1,
\label{t24}
\end{equation}%
and
\begin{equation}
\left\langle a^{\dagger }b^{\dagger }ab\right\rangle =\frac{%
N_{m+1,n+1}-N_{m+1,n}-N_{m,n+1}}{N_{m,n}}+1.  \label{t25}
\end{equation}%
Thus the cross-correlation function $g_{m,n}$ can be obtained by \cite{34}
\begin{eqnarray}
g_{m,n}(r) &=&\frac{\left\langle a^{\dagger }b^{\dagger }ab\right\rangle }{%
\left\langle a^{\dagger }a\right\rangle \left\langle b^{\dagger
}b\right\rangle }-1  \notag \\
&=&\frac{N_{m+1,n+1}N_{m,n}-N_{m,n+1}N_{m+1,n}}{\left(
N_{m,n+1}-N_{m,n}\right) \left( N_{m+1,n}-N_{m,n}\right) }.  \label{t26}
\end{eqnarray}%
The positivity of the cross-correlation function $g_{m,n}$ refers to
correlations between the two modes. In particular, when $m=n=0$
corresponding to the TMSTS, noticing $N_{0,0}=1,N_{0,1}=N_{1,0}=B_{1}$, and $%
N_{1,1}=\upsilon $, then Eq.(\ref{t26}) reduces to $g_{0,0}(r)=\left( 2\bar{n%
}+1\right) ^{2}\sinh ^{2}r\cosh ^{2}r/\left( B_{1}-1\right) ^{2},$which
implies that the parameter $g_{0,0}(r)$ is always positive for any $\bar{n}$
and non-zero squeezing ($B_{1}\neq 1$). Further, for the case of $\bar{n}=0,$
$g_{0,0}(r)=\coth ^{2}r,$ which is just the correlation function of the
TMSVS; while for $r=0,$ i.e., the TMSTS, $g_{0,0}(0)=0$, so there is no
correlation between two thermal states, as expected. On the other hand, when
$m=0,n=1$, noticing $N_{1,2}=B_{1}\left( 3\upsilon -\omega \right)
,N_{0,2}=2B_{1}^{2}$, and $P_{1}^{(0,1)}\left( x\right) =(3x-1)/2,$ then Eq.(%
\ref{t26}) becomes $g_{0,1}(r)=\left( \upsilon -\omega \right) B_{1}/[\left(
2B_{1}-1\right) \left( \upsilon -B_{1}\right) ]$. Noticing that $\upsilon
-B_{1}>0$ and $\left( 2B_{1}-1\right) >0$, and $\upsilon -\omega =\frac{1}{2}%
\left( 2\bar{n}+1\right) ^{2}\sinh ^{2}2r\geqslant 0,$ so $g_{0,1}(r)\ $is
always positive.

\begin{figure}[tbp]
\label{Fig1} \centering\includegraphics[width=8cm]{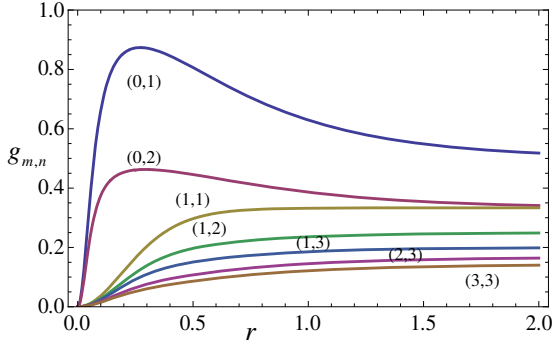}
\caption{{\protect\small (Color online) Cross-correlation function between
the two modes }${\protect\small a}${\protect\small \ and }${\protect\small b}
${\protect\small \ as a function of }${\protect\small r}${\protect\small \
for different parameters (m,n) and }${\protect\small \bar{n}=0.01.}$}
\end{figure}

In order to see clearly the variation of $g_{m,n}$-parameter, we plot the
graph of $g_{m,n}$ as the function of $r$ for some different ($m,n$) and $%
\bar{n}$ values. It is shown that $g_{m,n}$ are always larger than zero,
thus there exist correlations between the two modes. This implies that the
nonclassicality is enhanced by adding photon to squeezed state. For given ($%
m,n$) and $\bar{n}$ values, $g_{m,n}$ increases as $r$ increasing; while $%
g_{m,n}$ decreases as $\bar{n}$ decreasing for a given ($m,n$) value. It is
interesting to notice that for single-photon-addition TMSTS, the $g_{m,n}$
parameter presents its maximum value, which implies that
single-photon-addition TMSTS may possess a stronger nonclassicality than the
other TMSTSs. To compare the further nonclassicality of quantum states for a
different number added case, the measurments based on the volume of the
negative part of the Wigner function \cite{35}, on the nonclassical depth
\cite{36}, and on the entanglement potential \cite{37}, Vogel's
noncalssicality criterion \cite{38} and the Klyshko criterion \cite{39} may
be other alternative methods.

\subsection{Distribution of photon number of the PA-TMSTS}

In order to obtain the photon number distribution (PND) of the PA-TMSTS, we
begin with evaluating the PND of TMSTS. For two-mode case described by
density operator $\rho ^{S}$, the PND is defined by $\mathcal{P}%
(m_{a},n_{b})=\left\langle m_{a},n_{b}\right\vert \rho ^{S}\left\vert
m_{a},n_{b}\right\rangle .$ Employing the non-normalized coherent state $%
\left\vert \alpha \right\rangle =\exp [\alpha a^{\dag }]\left\vert
0\right\rangle $ leading to $\left\vert n\right\rangle =\frac{1}{\sqrt{n!}}%
\left. \frac{d^{n}}{d\alpha ^{n}}\left\vert \alpha \right\rangle \right\vert
_{\alpha =0}$ $\left( \left\langle \beta \right. \left\vert \alpha
\right\rangle =e^{\alpha \beta ^{\ast }}\right) $, as well as the normal
ordering form of $\rho ^{S}$ in Eq.(\ref{t7}), the probability of finding $%
\left( m_{a},n_{b}\right) $ photons in the two-mode field is given by%
\begin{eqnarray}
\mathcal{P}(m_{a},n_{b}) &=&\frac{A_{1}}{m_{a}!n_{b}!}\frac{d^{2m_{a}+2n_{b}}%
}{d\alpha ^{m_{a}}d\alpha ^{\ast m_{a}}d\beta ^{n_{b}}d\beta ^{\ast n_{b}}}
\notag \\
&&\times \left. e^{\left( 1-A_{3}\right) \left( \alpha ^{\ast }\alpha +\beta
^{\ast }\beta \right) +A_{2}\left( \alpha \beta +\alpha ^{\ast }\beta ^{\ast
}\right) }\right\vert _{\alpha ,\beta ,\alpha ^{\ast },\beta ^{\ast }=0}
\notag \\
&=&A_{1}\left[ \allowbreak \bar{n}\left( \bar{n}+1\right) \right]
^{n_{b}-m_{a}}\frac{\mu \allowbreak ^{m_{a}}}{\nu ^{n_{a}}}%
P_{m_{a}}^{(n_{b}-m_{a},0)}\left( \chi \right) ,  \label{t27}
\end{eqnarray}%
where in the last step, we have used the new formula in Eq.(\ref{t20}), and%
\begin{eqnarray}
\nu &=&\left( 2\bar{n}+\allowbreak 1\right) \cosh ^{2}r+\bar{n}^{2},  \notag
\\
\mu &=&\left( 2\bar{n}+1\right) \cosh ^{2}r-\left( \bar{n}+1\right) ^{2},
\notag \\
\chi &=&\frac{\left( \left( 2\bar{n}+1\right) \sinh 2r\right)
^{2}+4\allowbreak \bar{n}^{2}\left( \bar{n}+1\right) ^{2}}{\left( \left( 2%
\bar{n}+1\right) \sinh 2r\right) ^{2}-4\allowbreak \bar{n}^{2}\left( \bar{n}%
+1\right) ^{2}}.  \label{t28}
\end{eqnarray}%
Thus the PND of TMSTS is also related to Jacobi polynomials of the parameter
$\chi $. In particular, when $\bar{n}\rightarrow 0$ leading to $\chi
\rightarrow 1$, corresponding to the two-mode squeezed vacuum, Eq.(\ref{t27}%
) reduces to%
\begin{align}
& \mathcal{P}_{\bar{n}\rightarrow 0}(m_{a},n_{b})  \notag \\
& =\lim_{\bar{n}\rightarrow 0}\frac{m_{a}!n_{b}!}{\left[ \bar{n}^{2}+\left( 2%
\bar{n}+1\right) \cosh ^{2}r\right] ^{m_{a}+n_{b}+1}}  \notag \\
& \times \sum_{l=0}^{\min [m_{a},n_{a}]}\frac{\left( \allowbreak \allowbreak
2\bar{n}+1\right) ^{2l}\left[ \bar{n}\left( \bar{n}+1\right) \right]
^{m_{a}+n_{b}-2l}\sinh ^{2l}2r}{2^{2l}\left( l!\right) ^{2}\left(
n_{b}-l\right) !\left( m_{a}-l\right) !}  \notag \\
& =\frac{\tanh ^{2m_{a}}r}{\cosh ^{2}r}\delta _{m_{a},n_{b}},  \label{t29}
\end{align}%
which is just the PND of two-mode squeezed vacuum state \cite{23}. On the
other hand, when $r\rightarrow 0$ corresponding to the case of two-mode
thermal state, leading to $\chi \rightarrow -1,\nu \rightarrow \left( \bar{n}%
+1\right) ^{2},\mu \rightarrow -\bar{n}^{2},A_{1}\rightarrow 1/\left( \bar{n}%
+1\right) ^{2}$ and noting $P_{m_{a}}^{(n_{b}-m_{a},0)}(-1)=(-1)^{m_{a}},$
thus Eq.(\ref{t27}) becomes%
\begin{equation}
\mathcal{P}_{r\rightarrow 0}(m_{a},n_{b})=\allowbreak \frac{\bar{n}^{n_{b}}}{%
\left( \bar{n}+1\right) ^{n_{b}+1}}\frac{\bar{n}^{m_{a}}}{\left( \bar{n}%
+1\right) ^{m_{a}+1}},  \label{t30}
\end{equation}%
which is just the product of PNDs of two thermal fields, as expected.

Using the result (\ref{t27}) and noticing $a^{m}b^{n}\left\vert
m_{a},n_{b}\right\rangle =\sqrt{m_{a}!n_{b}!/(m_{a}-m)!(n_{b}-n)!}\left\vert
m_{a}-m,n_{b}-n\right\rangle $, we can directly obtain the PND $\mathcal{%
\bar{P}}^{SA}(m_{a},n_{b})\equiv \left\langle m_{a},n_{b}\right\vert \rho
^{SA}\left\vert m_{a},n_{b}\right\rangle $ of the PA-TMSTS as%
\begin{eqnarray}
&&\mathcal{\bar{P}}^{SA}(m_{a},n_{b})  \notag \\
&=&\frac{N_{m,n}^{-1}m_{a}!n_{b}!}{(m_{a}-m)!(n_{b}-n)!}  \notag \\
&&\times \left\langle m_{a}-m,n_{b}-n\right\vert \rho ^{S}\left\vert
m_{a}-m,n_{b}-n\right\rangle  \notag \\
&=&\frac{N_{m,n}^{-1}m_{a}!n_{b}!}{(m_{a}-m)!(n_{b}-n)!}\mathcal{P}%
(m_{a}-m,n_{b}-n).  \label{t31}
\end{eqnarray}%
Eq.(\ref{t31}) is a Jacobi polynomial with a condition $m_{a}\geqslant m$
and $n_{b}\geqslant n$ which shows that the photon-number ($m_{a},n_{b}$)
involved in PA-TMSTS are always no-less than the photon-number ($m,n$)
operated on the TMSTS, and there is no photon distribution when $m_{a}<m$
and $n_{b}<n$\textbf{. }Here we should point out that this result (\ref{t27}%
) can be applied directly to calculate the PND of some other non-Gaussian
states generated by subtracting photons from (or adding photons to) two-mode
squeezed thermal states, such as $a^{m}b^{n}\rho ^{S}a^{\dagger m}b^{\dagger
n},$ and $a^{m}b^{\dag n}\rho ^{S}b^{n}a^{\dag m}$.

\subsection{Antibunching effect of\textit{\ }the\textit{\ }PA-TMSTS}

Next we will discuss the antibunching for the PA-TMSTS. The criterion for
the existence of antibunching in two-mode radiation is given by \cite{40}%
\begin{equation}
R_{ab}\equiv \frac{\left\langle a^{\dagger 2}a^{2}\right\rangle
+\left\langle b^{\dagger 2}b^{2}\right\rangle }{2\left\langle a^{\dagger
}ab^{\dagger }b\right\rangle }-1<0.  \label{t32}
\end{equation}%
In a similar way to Eq.(\ref{t26}) we have%
\begin{eqnarray}
\left\langle a^{\dagger 2}a^{2}\right\rangle &=&\frac{N_{m+2,n}}{N_{m,n}}-4%
\frac{N_{m+1,n}}{N_{m,n}}+2,  \notag \\
\left\langle b^{\dagger 2}b^{2}\right\rangle &=&\frac{N_{m,n+2}}{N_{m,n}}-4%
\frac{N_{m,n+1}}{N_{m,n}}+2.  \label{t33}
\end{eqnarray}%
Thus, for the state $\rho ^{SA}$, substituting Eqs.(\ref{t33}), (\ref{t25})
and (\ref{t21}) into Eq.(\ref{t32}), yields%
\begin{eqnarray}
R_{ab} &=&\frac{N_{m+2,n}+N_{m,n+2}+2\left( \Omega -N_{m+1,n+1}\right) }{%
2\left( N_{m+1,n+1}+\Omega \right) },  \label{t34} \\
&&\left( \Omega =N_{m,n}-N_{m+1,n}-N_{m,n+1}\right) .  \notag
\end{eqnarray}%
In particular, when $m=n=0$ (corresponding to the TMSTS) leading to $%
N_{0,0}=1,N_{0,1}=N_{1,0}=B_{1}$, and $N_{1,1}=\upsilon $, $%
N_{0,2}=N_{2,0}=B_{1}^{2}$, thus Eq.(\ref{t34}) becomes
\begin{equation}
R_{ab,m=n=0}=-\frac{\left( 2\bar{n}+1\right) \left( 4\cosh 2r+\left( 2\bar{n}%
+1\right) \sinh ^{2}2r\right) }{\left( 2\bar{n}+1\right) \left[ \left( 2\bar{%
n}+1\right) \cosh 4r-2\cosh 2r\right] +1}.  \label{t35}
\end{equation}%
From Eq.(\ref{t35}), it is easily seen that $R_{ab,m=n=0}<0$ for any $\bar{n}
$ and non-zero $r$ values. In addition, when $m=n,$ the PA-TMSTS\textit{\ }%
can always be antibunching for a small value $\bar{n}$ (see Fig.2(a)).
However, for any parameter values $m,n(m\neq n)$, the case\textit{\ }is not
true. The $R_{ab}$ parameter as a function of $r$ and $m,n$ is plotted in
Fig. 2. It is easy to see that, for a given $m$ the PA-TMSTS\textit{\ }%
presents the antibunching effect when the squeezing parameter $r$ exceeds to
a certain threshold value. For instance, when $m=0\ $and $n=1$ then $R_{ab}\
$may be less than zero with $r>0.1$ thereabout ($\bar{n}=0.01$). The value $%
R_{ab}$ parameter increases with $\bar{n}$ increasing.

\begin{figure}[tbp]
\label{Fig2} \centering\includegraphics[width=8cm]{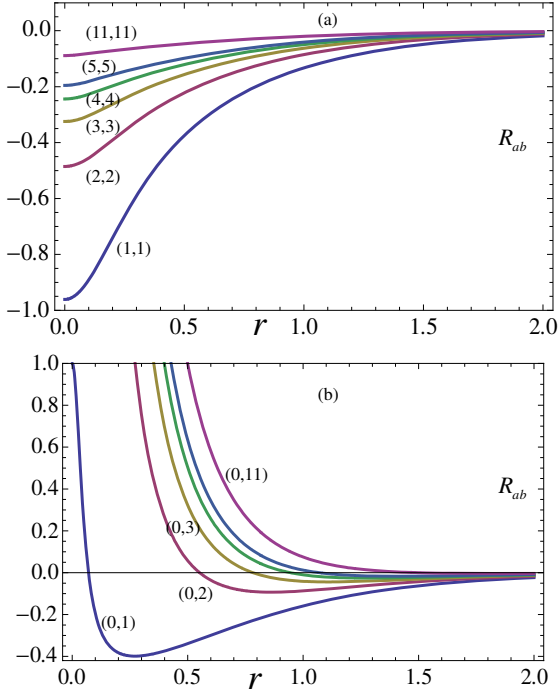}
\caption{{\protect\small (Color online) }${\protect\small R}_{ab}$%
{\protect\small \ as a function of }${\protect\small r}${\protect\small \
for different parameters (m,n) and }${\protect\small \bar{n}=0.01.}$}
\end{figure}

\subsection{Wigner function of\textit{\ }PA-TMSTS}

For further discussing the nonclassicality of PA-TMSTS, we examine its
Wigner function (WF) whose partial negativity implies the highly
nonclassical properties of quantum states. In this section, we derive the
analytical expression of WF for the PA-TMSTS. The normally ordering form of
the PA-TMSTS shall be used to realize our purpose.

For the two-mode case, the WF $W\left( \alpha ,\beta \right) $ associated
with a quantum state $\rho $ can be derived as follows \cite{42}:%
\begin{eqnarray}
W\left( \alpha ,\beta \right) &=&e^{2(\left\vert \alpha \right\vert
^{2}+\left\vert \beta \right\vert ^{2})}\int \frac{\mathtt{d}^{2}z_{1}%
\mathtt{d}^{2}z_{2}}{\pi ^{4}}\left\langle -z_{1},-z_{2}\right\vert \rho
\left\vert z_{1},z_{2}\right\rangle  \notag \\
&&\times \exp \left[ 2\left( \alpha z_{1}^{\ast }-\alpha ^{\ast
}z_{1}\right) +2\left( \beta z_{2}^{\ast }-\beta ^{\ast }z_{2}\right) \right]
,  \label{t36}
\end{eqnarray}%
where $\left\vert z_{1},z_{2}\right\rangle =\left\vert z_{1}\right\rangle
\left\vert z_{2}\right\rangle $ is the two-mode coherent state.

Substituting Eq.(\ref{t17}) into Eq.(\ref{t36}), we can finally obtain the
WF of the PA-TMSTS (see Appendix C),%
\begin{equation}
W_{m,n}\left( \alpha ,\beta \right) =W_{0}\left( \alpha ,\beta \right)
F_{m,n}\left( \alpha ,\beta \right) ,  \label{t37}
\end{equation}%
where $W_{0}\left( \alpha ,\beta \right) $ is the WF of TMSTS,
\begin{eqnarray}
W_{0}\left( \alpha ,\beta \right) &=&\frac{\pi ^{-2}}{\left( 2\bar{n}%
+1\right) ^{2}}\exp \left\{ -2\frac{\cosh 2r}{2\bar{n}+1}(\left\vert \alpha
\right\vert ^{2}+\left\vert \beta \right\vert ^{2})\right.  \notag \\
&&+\allowbreak \left. 2\frac{\sinh 2r}{2\bar{n}+1}\left( \beta \alpha
+\alpha ^{\ast }\beta ^{\ast }\right) \right\} ,  \label{t38}
\end{eqnarray}%
and
\begin{eqnarray}
F_{m,n}\left( \alpha ,\beta \right) &=&\frac{K_{3}^{m+n}}{N_{m,n}}%
\sum_{l=0}^{m}\sum_{j=0}^{n}\frac{\left( m!n!\right) ^{2}\left(
-K_{1}/K_{3}\right) ^{l+j}}{l!j!\left[ \left( m-l\right) !\left( n-j\right) !%
\right] ^{2}}  \notag \\
&&\times \left\vert H_{m-l,n-j}\left( \frac{R_{1}}{i\sqrt{K_{3}}},\frac{R_{3}%
}{i\sqrt{K_{3}}}\right) \right\vert ^{2},  \label{t39}
\end{eqnarray}%
where we have set%
\begin{eqnarray}
R_{1} &=&2\left( K_{1}\allowbreak \alpha -K_{3}\beta ^{\ast }\right)
,R_{3}=2\left( K_{1}\beta -K_{3}\alpha ^{\ast }\right) ,  \notag \\
K_{1} &=&\frac{\bar{n}+\cosh ^{2}r}{2\bar{n}+1},K_{3}=\frac{\sinh r\cosh r}{2%
\bar{n}+1}.  \label{t40}
\end{eqnarray}%
Equation (\ref{t37}) is just the analytical expression of the WF for the
PA-TMSTS, a real function as expected. It is obvious that the WF lost its
Gaussian property in phase space due to the presence of two-variable Hermite
polynomials $H_{m-l,n-j}\left( x,y\right) $.

From Eq.(\ref{t39}), we see that when $m=n=0$ corresponding to the TMSTS, $%
F_{0,0}=1,$ and $W_{m,n}\left( \alpha ,\beta \right) =W_{0}\left( \alpha
,\beta \right) $; whereas for the case of $m=0$ and $n\neq 0$, noticing $%
H_{0,n}\left( x,y\right) =y^{n}$ and $N_{0,n}=n!B_{1}^{n},$ Eq.(\ref{t37})
reduces to%
\begin{equation}
W_{0,n}\left( \alpha ,\beta \right) =W_{0}\left( \alpha ,\beta \right)
\left( -K_{1}/B_{1}\right) ^{n}L_{n}(\left\vert R_{3}\right\vert ^{2}/K_{1}),
\label{t41}
\end{equation}%
where $L_{n}$ is the $n$-order Laguerre polynomial. Eq. (\ref{t41}) is just
the WF of the PA-TMSTS generated by single-mode photon addition, which
becomes the WF of the negative binomial state $S(r)\left\vert
0,n\right\rangle $ with $\bar{n}=0$ [JOSAB, ??]. In particular, for the case
of single photon-addition, $n=1$, it is found that
\begin{equation}
W_{0,1}\left( 0,0\right) =-\frac{\left( \bar{n}+\cosh ^{2}r\right) /\left( 2%
\bar{n}+1\right) ^{3}}{\left( \cosh ^{2}r+\bar{n}\cosh 2r\right) \pi ^{2}},
\label{t42}
\end{equation}%
which implies that the WF of single PA-TMSTS always has its negative region
at the phase space center $\alpha =\beta =0$. The maximum value of \ $%
\left\vert W_{0,1}\right\vert $ decreases with the increasement of $\bar{n}$
and $r$ but not disappears, which can be seen clearly from Fig3,4. (a) and
(b). Further, there are more visible negative region than the WF for the
case of $m=n=1$. And the negative region will be absence for the latter with
the increasing $\bar{n}$ value (see Fig3,4. (c) and (d)). In addition, from
Figs 3 and 4, the squeezing in one of quadratures is clear, which can be
seen as an evidence of nonclassicality of the state. For a given value $m$\
and several different values $n$ ($\neq m$), the WF distributions are
presented in Fig.5, from which it is interesting to notice that there are
around $\left\vert m-n\right\vert $ wave valleys and $\left\vert
m-n\right\vert +1$ wave peaks.$\allowbreak $

\begin{figure}[tbp]
\label{Fig3} \centering\includegraphics[width=9cm]{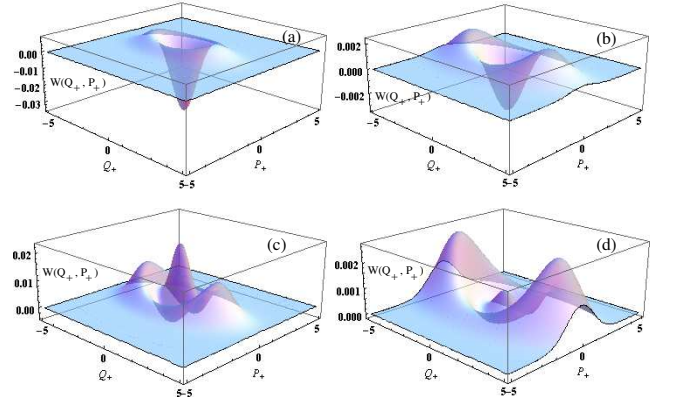}
\caption{{\protect\small (Color online) The Wigner function W(}$\protect%
\alpha ,\protect\beta ${\protect\small ) in phase space (}$Q_{+},P_{+}$%
{\protect\small ) for several different parameter values }$\left( m,n\right)
${\protect\small \ and }$\bar{n}${\protect\small \ with }$r=0.3.$%
{\protect\small \ (a) m=0,n=1,}$\bar{n}=0.2${\protect\small ; (b) m=0,n=1,}$%
\bar{n}=1${\protect\small ; (c) m=n=1,}$\bar{n}=0.2${\protect\small \ and
(d) m=n=1,}$\bar{n}=1.$}
\end{figure}

\begin{figure}[tbp]
\label{Fig4} \centering\includegraphics[width=9cm]{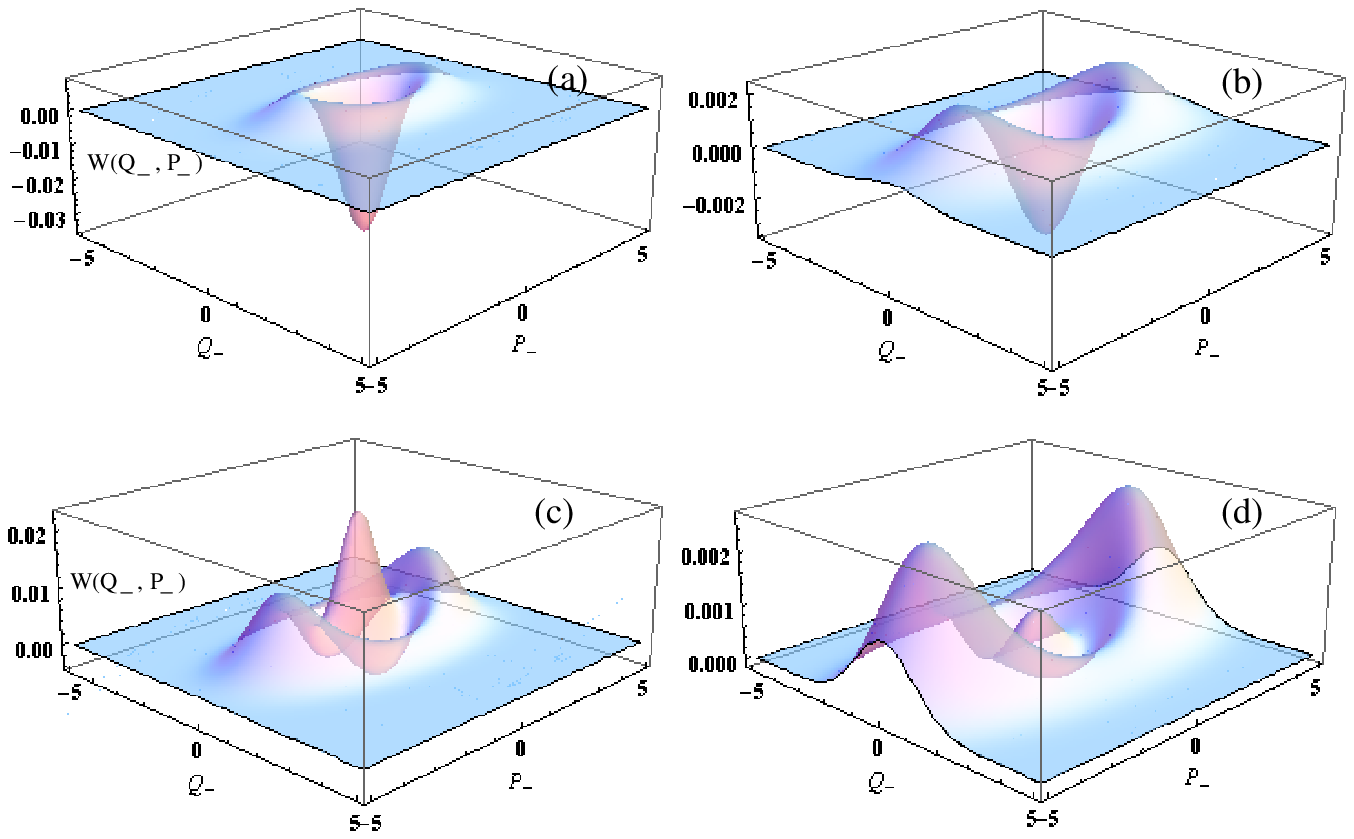}
\caption{{\protect\small (Color online) The Wigner function W(}$\protect%
\alpha ,\protect\beta ${\protect\small ) in phase space (}$Q_{-},P_{-}$%
{\protect\small ) for several different parameter values }$\left( m,n\right)
${\protect\small \ and }$\bar{n}${\protect\small \ with }$r=0.3.$%
{\protect\small \ (a) m=0,n=1,}$\bar{n}=0.2${\protect\small ; (b) m=0,n=1,}$%
\bar{n}=1${\protect\small ; (c) m=n=1,}$\bar{n}=0.2${\protect\small \ and
(d) m=n=1,}$\bar{n}=1.$}
\end{figure}

\begin{figure}[tbp]
\label{Fig5} \centering\includegraphics[width=9cm]{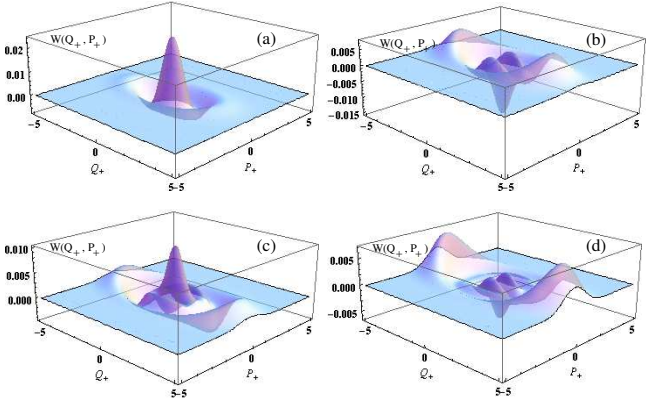}
\caption{{\protect\small (Color online) The Wigner function W(}$\protect%
\alpha ,\protect\beta ${\protect\small ) in phase space (}$Q_{+},P_{+}$%
{\protect\small ) for several different parameter values }$\left( m,n\right)
${\protect\small \ with }$\bar{n}=0.2$ {\protect\small and \ }$r=0.3.$%
{\protect\small \ (a) m=0,n=2; (b) m=1,n=2; (c) m=1,n=3, and (d) m=2,n=3}$.$}
\end{figure}

\section{Entanglement properties of the PA-TMSTS}

It is well known that photon subtraction/addition can be applied to improve
entanglement between Gaussian states \cite{14,43}, loophole-free tests of
Bell's inequality \cite{44}, and quantum computing \cite{18}. In this
section, we examine the entanglement properties of PA-TMSTS only with single
and two photon-addition. Here, for a bipartite continuous variable state, we
shall take the Shchukin-Vogel (SV) \cite{38} criteria to describe the
inseparability of PA-TMSTS.

According to the SV criteria, the sufficient condition of inseparability is
\begin{equation}
SV_{m,n}\equiv \left\langle a^{\dag }a-\frac{1}{2}\right\rangle \left\langle
b^{\dag }b-\frac{1}{2}\right\rangle -\left\langle a^{\dag }b^{\dag
}\right\rangle \left\langle ab\right\rangle <0.  \label{t43}
\end{equation}%
In a similar way to derive the normalization factor Eq.(\ref{t21}),\ using
Eqs.(\ref{t17}) and (\ref{t18}), we have%
\begin{equation}
\left\langle a^{\dag }b^{\dag }\right\rangle =\frac{N_{m,m+1,n,n+1}}{N_{m,n}}%
,\text{ }\left\langle ab\right\rangle =\frac{N_{m+1,m,n+1,n}}{N_{m,n}},
\label{t44}
\end{equation}%
where we have set%
\begin{eqnarray}
N_{l,p,q,r} &\equiv &\left. \frac{\partial ^{l+p+q+r}}{\partial \tau
^{l}\partial t^{p}\partial \tau ^{\prime q}\partial t^{\prime r}}e^{\left(
\tau ^{\prime }t^{\prime }+\tau t\right) B_{1}+\left( \tau \tau ^{\prime
}+tt^{\prime }\right) B_{2}}\right\vert _{t,\tau ,t^{\prime },\tau ^{\prime
}=0}  \notag \\
&=&\sum_{s=0}^{{}}\frac{l!r!q!p!B_{1}^{l-q+2r}B_{2}^{q-r}\left(
B_{2}^{2}/B_{1}^{2}\right) ^{s}\delta _{p+q,l+r}}{s!\left( q-r+s\right)
!\left( r-s\right) !\left( l+r-q-s\right) !}.  \label{t45}
\end{eqnarray}%
Thus $SV$ is given by%
\begin{eqnarray}
SV_{m,n} &=&\left( \frac{N_{m+1,n}}{N_{m,n}}-\frac{3}{2}\right) \left( \frac{%
N_{m,n+1}}{N_{m,n}}-\frac{3}{2}\right)  \notag \\
&&-\frac{N_{m,m+1,n,n+1}}{N_{m,n}}\frac{N_{m+1,m,n+1,n}}{N_{m,n}}.
\label{t46}
\end{eqnarray}%
Next, we examine two special cases. For the case of $m=0,n=1$, using Eqs.(%
\ref{t21}) and (\ref{t45}), as well as noticing $%
N_{0,1,1,2}=N_{1,0,2,1}=2B_{1}B_{2}$, Eq.(\ref{t46}) becomes%
\begin{equation}
SV_{0,1}=\left( \frac{\upsilon }{B_{1}}-\frac{3}{2}\right) \left( 2B_{1}-%
\frac{3}{2}\right) -4B_{2}^{2}.  \label{t47}
\end{equation}%
While for the case of $m=n=1$, it is shown that ($N_{1,2,1,2}=N_{2,1,2,1}=2%
\left( \allowbreak 2B_{1}^{2}+B_{2}^{2}\right) B_{2}$)%
\begin{equation}
SV_{1,1}=\left( B_{1}\left( 3-\frac{\omega }{\upsilon }\right) -\frac{3}{2}%
\right) ^{2}-4\frac{\left( \allowbreak 2B_{1}^{2}+B_{2}^{2}\right)
^{2}B_{2}^{2}}{\upsilon ^{2}}.  \label{t48}
\end{equation}%
In particular, when $\bar{n}=0$, i.e., the single PA-TMSVS, Eq.(\ref{t47})
is always negative for any $r>0,$ as expected (also see Fig.6 (a)). In
general, it is difficult to obtain the explicit expressions of the
sufficient condition of inseparability for the above cases. Here, we appeal
to the number calculation shown in Fig.6. It is shown that for single
PA-TMSTS with a smaller average photon number $\bar{n}$, the condition $%
SV_{0,1}<0$ can always be satisfied only if $r>0$; while for a larger $\bar{n%
}$ then the condition $SV_{0,1}<0$ is satisfied only when the squeezing
parameter $r$ exceeds a certain threshold value $r_{a}$. However, it is very
interesting to notice that for the photon-subtraction TMSTS, there is a
threshold value $r_{c}$ for any $\bar{n}$, i.e., $r>r_{c}\equiv \frac{1}{2}%
\ln (2\bar{n}+1)$ \cite{45}, which is different from the case of single
PA-TMSTS. For instance, for $\bar{n}=1,$ the two threshold values are $%
r_{a}\approx 0.31$ and $r_{c}\approx 0.55$. This comparision may imply that
the photon-addition to the TMSTS can be more effective for the entanglement
enhancement than the photon-subtraction from the TMSTS. On the other hand,
for the case of the PA-TMSTS with $m=n=1$ (see Fig.6 (b)), it is found that
a certain threshold is needed for satisfying this condition $SV_{1,1}<0,$
which is also smaller than that of the photon-subtraction TMSTS.

\begin{figure}[tbp]
\label{Fig6} \centering\includegraphics[width=9cm]{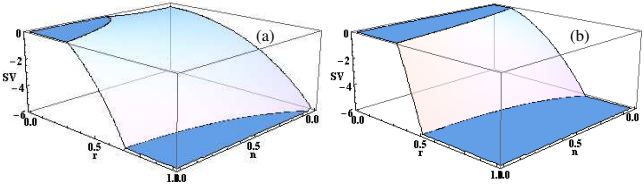}
\caption{{\protect\small (Color online) The sufficient condition of
inseparability as the function of }${\protect\small r}${\protect\small \ and }%
$\bar{n},$ {\protect\small for (a) m=0 and n=1}; {\protect\small (b) m=n=1.}}
\end{figure}

\section{Quantum teleportation with PA-TMSTS}

As mentioned above, photon-subtraction from or photon-addition to bipartite
Gaussian states can be used to improve the entanglement.\ In this section,
we investigate the quantum teleportation with PA-TMSTS, especially for the
cases $m=0,n=1$ and $m=n=1$. The role of teleportation in the CV quantum
information is analyzed in the review Ref.\cite{46}.

Here, we consider the QT by using PA-TMSTS as entangled resource. Using the
normal ordering form Eq.(\ref{t17}) and noticing the displacement operator $%
D_{a}\left( \alpha \right) =e^{\left\vert \alpha \right\vert
^{2}/2}e^{-\alpha ^{\ast }a}e^{\alpha a^{\dagger }}$, then the
characteristic function (CF) of PA-TMSTS is given by (see Appendix D)%
\begin{eqnarray}
\chi _{E}\left( \alpha ,\beta \right) &=&\frac{1}{N_{m,n}}e^{-(B_{1}-\frac{1%
}{2})(\left\vert \alpha \right\vert ^{2}+\left\vert \beta \right\vert
^{2})+B_{2}\left( \alpha \beta +\alpha ^{\ast }\beta ^{\ast }\right) }
\notag \\
&&\times \frac{\partial ^{2m+2n}}{\partial \tau ^{m}\partial \tau ^{\prime
n}\partial t^{m}\partial t^{\prime n}}e^{B_{1}\left( \tau ^{\prime
}t^{\prime }+t\tau \right) +B_{2}\left( t^{\prime }t+\tau ^{\prime }\tau
\right) }  \notag \\
&&\times e^{t\left( \alpha B_{1}-\allowbreak B_{2}\beta ^{\ast }\right)
+\tau \left( \beta B_{2}-B_{1}\alpha ^{\ast }\right) }  \notag \\
&&\times \left. e^{\tau ^{\prime }\left( \alpha B_{2}-B_{1}\beta ^{\ast
}\right) +t^{\prime }\left( \beta \allowbreak B_{1}-B_{2}\alpha ^{\ast
}\right) }\right\vert _{t,\tau ,t^{\prime },\tau ^{\prime }=0},  \label{t49}
\end{eqnarray}%
where for further calculation the differential form of $\chi _{E}$ is kept.

To quantify the performance of a QT protocol, the fidelity of QT is commonly
used as a measure, $\mathcal{F=}\mathtt{tr}\left( \rho _{in}\rho
_{out}\right) $, a overlap between a pure input state $\rho _{in}$ and the
output (teleported, mixed) state $\rho _{out}$. For a CV system, a
teleportation protocol has been given in terms of the CFs of the quantum
states involved (input, source and teleported (output) states) \cite{47}. It
is shown that the CF $\chi _{out}\left( \eta \right) $ of the output state
has a remarkably factorized form%
\begin{equation}
\chi _{out}\left( \eta \right) =\chi _{in}\left( \eta \right) \chi
_{E}\left( \eta ^{\ast },\eta \right) ,  \label{t50}
\end{equation}%
where $\chi _{in}\left( \eta \right) $ and $\chi _{E}\left( \eta ^{\ast
},\eta \right) $ are the CFs of the input state and the entangled source,
respectively. Then the fidelity of QT of CVs reads \cite{47}%
\begin{equation}
\mathcal{F=}\int \frac{d^{2}\eta }{\pi }\chi _{in}\left( \eta \right) \chi
_{out}\left( -\eta \right) .  \label{t51}
\end{equation}%
Here, we consider the Braunstein and Kimble protocol \cite{48} of QT for
single-mode coherent-input states $\left\vert \gamma \right\rangle $. Note
that the fidelity is independent of amplitude of the coherent state, thus
for simplicity we take $\gamma =0$, then we have only to calculate the
fidelity of the vacuum input state with the CF $\chi _{in}\left( \eta
\right) =\exp [-\left\vert \eta \right\vert ^{2}/2]$. On substituting these
CFs into Eq.(\ref{t51}), we worked out the fidelity for teleporting a
coherent state by using the PA-TMSTS as an entangled resource,%
\begin{eqnarray}
\mathcal{F}_{m,n}^{\bar{n}} &=&\frac{(m+n)!}{B_{1}-B_{2}}\frac{\left(
B_{1}+B_{2}\right) ^{m+n}}{2^{m+n+1}N_{m,n}}  \notag \\
&=&\frac{\left[ \allowbreak \left( 2\bar{n}+1\right) e^{2r}+\allowbreak 1%
\right] ^{m+n}}{\allowbreak \left( 2\bar{n}+1\right) e^{-2r}+1}\frac{(m+n)!}{%
2^{2m+2n}N_{m,n}}.  \label{t52}
\end{eqnarray}%
It can be seen that the fidelity is not only dependent on the parameter $r$,
the average photon-number $\bar{n}$, but also on the photon number $\left(
m,n\right) $ added to each mode of the TMSTS. In particular, when $m=n=0,$
Eq.(\ref{t52}) just reduces to%
\begin{equation}
\mathcal{F}_{0,0}^{\bar{n}}=\frac{1}{\allowbreak \left( 2\bar{n}+1\right)
e^{-2r}+1},  \label{t53}
\end{equation}%
which leads to the condition $r>\frac{1}{2}\ln \left( 2\bar{n}+1\right) $
for satisfying the effective QT with $\mathcal{F}>\frac{1}{2}$ which is the
classical limit. In addition, for the case of $\bar{n}=0$, i.e., the
photon-added TMSVS, Eq.(\ref{t52}) becomes
\begin{equation}
\mathcal{F}_{m,n}^{0}=\frac{\left( \allowbreak e^{2r}+\allowbreak 1\right)
^{n+m}}{\allowbreak e^{-2r}+1}\frac{(m+n)!}{2^{2m+2n}N_{m,n}}.  \label{t54}
\end{equation}%
Further,$\ $when $\left( m,n\right) =\left( 0,0\right) ,\left( 1,1\right) $
and $\left( 0,1\right) $, Eq.(\ref{t53}) just reduce, respectively, to%
\begin{eqnarray}
\mathcal{F}_{0,0}^{0} &\mathcal{=}&(1+\tanh r)/2,  \notag \\
\mathcal{F}_{1,1}^{0} &=&\frac{(1+\tanh r)^{3}}{4(1+\tanh ^{2}r)},  \notag \\
\mathcal{F}_{0,1}^{0} &=&\frac{1+\tanh r}{4\left( 1-\tanh r\right) }\text{%
sech}^{2}r.  \label{t55}
\end{eqnarray}%
The two expressions $\mathcal{F}_{0,0}^{0}$ and $\mathcal{F}_{1,1}^{0}$ are
agreement with Eqs.(15) and (17) in Ref. \cite{49}.

In Fig. 7, for some given $\left( m,n\right) $ values, the fidelity of
teleporting the coherent state is shown as a function of $r$ by using the
PA-TMSTS as the entangled resource. It is shown that the fidelity with this
resource is smaller than that with TMSTS, although the PA-TMSTS posesses
larger entanglement \cite{49}. In addition, for the symmetrical case $m=n$,
when the squeezing parameter $r$ exceeds a certain threshold value, the
fidelity increases with a increasing $m$ (see Fig.7(c)); while for
non-symmetric case $m\neq n$, the fidelity decreases with increasing $n\ $%
(see Fig.7(b)). For the former, the threshold value $r$ decreases with
increasing $m\left( =n\right) $; the case is not true for the latter. This
indicate that the symmetrical PA-TMSTS may be more effective for QT than the
non-symmetric case.

\begin{figure}[tbp]
\label{Fig7} \centering\includegraphics[width=9cm]{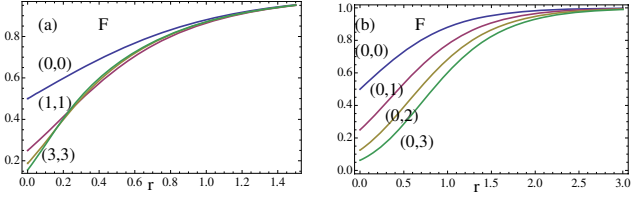}
\caption{{\protect\small (Color online) The fidelity as the function of }$%
{\protect\small r}${\protect\small \ for several different (m,n) values and }%
$\bar{n}{\protect\small =0.01}.$}
\end{figure}

\section{Conclusions}

In this paper, we introduce the PA-TMSTS and investigate its entanglement
and nonclassicality. By using the coherent state representation of thermal
state, the normally and antinormally ordering forms of the TMSTS are
obtained. Based on this, the normalization factor of the PA-TMSTS is
derived, which is related to the Jacobi polynomials of the squeezing
parameter $r$ and average photon number $\bar{n}$ of the thermal state. Then
we discuss the nonclassical properties by using cross-correlation function,
distribution of photon number, antibunching effect and the negativity of its
WF. It is found that the WF lost its Gaussian property in phase space due to
the presence of two-variable Hermite polynomials and the WF of single
PA-TMSTS always has its negative region at the center of phase space.
Further, there are more visible negative region than the WF for the case of $%
m=n=1$. And the negative region will be absence for the latter with the
increasing $\bar{n}$ value. The entanglement properties of the PA-TMSTS by
Shchukin-Vogel criteria and the quantum teleportation. It is shown that the
photon-addtion to the TMSTS can be more effective for the entanglement
enhancement than the photon-subtraction from the TMSTS; And using the
PA-TMSTS as an entangled resource, the fidelity for teleporting a coherent
state is not only dependent on the parameter $r$, the average photon-number $%
\bar{n}$, but also on the photon number $\left( m,n\right) $ added to each
mode of the TMSTS. From this point, the symmetrical PA-TMSTS may be more
effective for quantum teleportation than the non-symmetric case.

\textbf{Acknowledgments: } This work was supported by the NSFC (Grant No.
60978009), the Major Research Plan of the NSFC (Grant No. 91121023), and the \textquotedblleft
973\textquotedblright\ Project (Grant No. 2011CBA00200), and the Natural
Science Foundation of Jiangxi Province of China (No. 2010GQW0027) as well as
the Sponsored Program for Cultivating Youths of Outstanding Ability in
Jiangxi Normal University.

\textbf{Appendix A: Derivation of Eq.(\ref{t18})}

According to the normalization condition, $\mathtt{tr}\rho ^{SA}=1,$ we have%
\begin{align}
N_{m,n}& =A_{1}\mathtt{tr}\left[ \colon a^{\dagger m}b^{\dagger
n}e^{A_{2}\left( a^{\dagger }b^{\dagger }+ab\right) -A_{3}\left( a^{\dagger
}a+b^{\dagger }b\right) }a^{m}b^{n}\colon \right]  \notag \\
& =\frac{A_{1}\partial ^{2m+2n}}{\partial \tau ^{m}\partial t^{m}\partial
\tau ^{\prime n}\partial t^{\prime n}}\mathtt{tr}\left[ \colon
e^{A_{2}\left( a^{\dagger }b^{\dagger }+ab\right) -A_{3}\left( a^{\dagger
}a+b^{\dagger }b\right) +\tau a^{\dagger }+ta+\tau ^{\prime }b^{\dagger
}+t^{\prime }b}\colon \right] .  \tag{A1}
\end{align}

Using the completeness relation of coherent state $\int d^{2}\alpha
d^{2}\beta \left\vert \alpha ,\beta \right\rangle \left\langle \alpha ,\beta
\right\vert /\pi ^{2}=1$ and Eq.(\ref{t9}), Eq.(A1)%
\begin{align}
N_{m,n}& =A_{1}\int \frac{d^{2}\alpha d^{2}\beta }{\pi ^{2}}\left\vert
\alpha \right\vert ^{2m}\left\vert \beta \right\vert ^{2n}e^{A_{2}\left(
\alpha ^{\ast }\beta ^{\ast }+\alpha \beta \right) -A_{3}(\left\vert \alpha
\right\vert ^{2}+\left\vert \beta \right\vert ^{2})}  \notag \\
& =\frac{A_{1}\partial ^{2m+2n}}{\partial \tau ^{m}\partial t^{m}\partial
\tau ^{\prime n}\partial t^{\prime n}}\int \frac{d^{2}\alpha d^{2}\beta }{%
\pi ^{2}}\exp \left[ -A_{3}(\left\vert \alpha \right\vert ^{2}+\left\vert
\beta \right\vert ^{2})\right.  \notag \\
& \left. +\left( A_{2}\beta +\tau \right) \alpha +\left( A_{2}\beta ^{\ast
}+t\right) \alpha ^{\ast }+\tau ^{\prime }\beta +t^{\prime }\beta ^{\ast }
\right] _{t,\tau ,t^{\prime },\tau ^{\prime }=0}.  \tag{A2}
\end{align}%
Using Eq.(\ref{t9}), (A2) becomes
\begin{align}
N_{m,n}& =\frac{A_{1}\partial ^{2m+2n}}{\partial \tau ^{m}\partial
t^{m}\partial \tau ^{\prime n}\partial t^{\prime n}}\frac{1}{A_{3}}\int
\frac{d^{2}\beta }{\pi }\exp \left[ -\frac{A_{3}^{2}-A_{2}^{2}}{A_{3}}%
\left\vert \beta \right\vert ^{2}\right.  \notag \\
& +\left. \left( \frac{A_{2}t}{A_{3}}+\tau ^{\prime }\right) \beta +\left(
\frac{A_{2}\tau }{A_{3}}+t^{\prime }\right) \beta ^{\ast }+\frac{\tau t}{%
A_{3}}\right] _{t,\tau ,t^{\prime },\tau ^{\prime }=0}  \notag \\
& =\text{LHS of Eq.(\ref{t18})},  \tag{A3}
\end{align}%
where we have used $A_{1}/(A_{3}^{2}-A_{2}^{2})=1$ and $B_{2}=A_{2}%
\allowbreak /(A_{3}^{2}-A_{2}^{2})=\left( 2\bar{n}+1\right) \sinh r\cosh r$
as well as $B_{1}=A_{3}/(A_{3}^{2}-A_{2}^{2})$.

\bigskip

\textbf{Appendix B: New expression of generating function for Jacobi
polynomials}

In this appendix, we shall prove Eq.(\ref{t20}). Rewriting%
\begin{equation}
H\equiv \left. \frac{\partial ^{2m+2n}}{\partial \tau ^{m}\partial
t^{m}\partial \tau ^{\prime n}\partial t^{\prime n}}e^{A\left( \tau ^{\prime
}t^{\prime }+\tau t\right) +B\left( \tau \tau ^{\prime }+t^{\prime }t\right)
}\right\vert _{t,\tau ,t^{\prime },\tau ^{\prime }=0}.  \tag{B1}
\end{equation}%
Expanding the exponential items, we see%
\begin{align}
H& =\sum_{l,j,k,s=0}^{\infty }\frac{A^{l+j}B^{s+k}}{l!j!k!s!}\left. \frac{%
\partial ^{2m+2n}}{\partial \tau ^{m}\partial t^{m}\partial \tau ^{\prime
n}\partial t^{\prime n}}\tau ^{k+j}t^{j+s}\tau ^{\prime l+k}t^{\prime
l+s}\right\vert _{t,\tau ,t^{\prime },\tau ^{\prime }=0}  \notag \\
& =\sum_{s=0}^{\min \left[ m,n\right] }\frac{\left( m!n!\right) ^{2}A^{n+m}}{%
s!s!\left( n-s\right) !\left( m-s\right) !}\left( \frac{B^{2}}{A^{2}}\right)
^{s}.  \tag{B2}
\end{align}%
Comparing Eq.(B2) with the standard expression of Jacobi polynomials,
\begin{equation}
P_{m}^{(\alpha ,\beta )}(x)=\left( \frac{x-1}{2}\right)
^{m}\sum_{k=0}^{m}\left(
\begin{array}{c}
m+\alpha \\
k%
\end{array}%
\right) \left(
\begin{array}{c}
m+\beta \\
m-k%
\end{array}%
\right) \left( \frac{x+1}{x-1}\right) ^{k},  \tag{B3}
\end{equation}%
we can find that taking $m\leqslant n$ and $y=\left( B^{2}+A^{2}\right)
/\left( B^{2}-A^{2}\right) $,
\begin{align}
H& =\sum_{s=0}^{m}\frac{\left( m!n!\right) ^{2}A^{n+m}}{s!s!\left(
n-s\right) !\left( m-s\right) !}\left( \frac{B^{2}}{A^{2}}\right) ^{s}
\notag \\
& =m!n!\left( \frac{y-1}{2}\right) ^{-m}A^{m+n}\left\{ \left( \frac{y-1}{2}%
\right) ^{m}\right.  \notag \\
& \times \left. \sum_{k=0}^{\min [m,n]}\frac{m!n!}{k!k!\left( n-k\right)
!\left( m-k\right) !}\left( \frac{y+1}{y-1}\right) ^{k}\right\}  \notag \\
& =m!n!A^{n-m}\left( B^{2}-A^{2}\right) ^{m}P_{m}^{(n-m,0)}\left( y\right) .
\tag{B4}
\end{align}%
In a similar way, for $n\leqslant m$, we also have%
\begin{equation}
H=m!n!A^{m-n}\left( B^{2}-A^{2}\right) ^{n}P_{n}^{(m-n,0)}\left( y\right) .
\tag{B5}
\end{equation}%
Thus we finish the proof of Eq.(\ref{t20}).

In addition, when $m=n$, Eq.(\ref{t20}) becomes%
\begin{align}
& \left. \frac{\partial ^{4m}}{\partial \tau ^{m}\partial t^{m}\partial \tau
^{\prime m}\partial t^{\prime m}}e^{A\left( \tau ^{\prime }t^{\prime }+\tau
t\right) +B\left( \tau \tau ^{\prime }+t^{\prime }t\right) }\right\vert
_{t,\tau ,t^{\prime },\tau ^{\prime }=0}  \notag \\
& =\left. \frac{\partial ^{4m}}{\partial \tau ^{m}\partial t^{m}\partial
\tau ^{\prime m}\partial t^{\prime m}}e^{\left( At^{\prime }+B\tau \right)
\tau ^{\prime }+\left( A\tau +Bt^{\prime }\right) t}\right\vert _{t,\tau
,t^{\prime },\tau ^{\prime }=0}  \notag \\
& =\left. \frac{\partial ^{4m}}{\partial \tau ^{m}\partial t^{m}}\left[
\left( At+B\tau \right) \left( A\tau +Bt\right) \right] ^{m}\right\vert
_{\tau ,t^{\prime }=0}  \notag \\
& =\left( m!\right) ^{2}\left( B^{2}-A^{2}\right) ^{m}P_{m}\left( \frac{%
B^{2}+A^{2}}{B^{2}-A^{2}}\right) ,  \tag{B6}
\end{align}%
where $P_{m}\left( x\right) $ is the $m$th Legendre polynomials. Eq.(B6) is
just a new formula for the generating function of Legendre polynomials $%
P_{m}(x)$, which is different from the new form found in Ref.\cite{50}. In
fact, one can check Eq. (B6) by expanding directly the whole exponential
items and comparing with the standard expression of Legendre polynomials.

\textbf{Appendix C: Derivation of Eq.(\ref{t37})}

Substituting Eq.(\ref{t17}) into Eq.(\ref{t36}) and usiing Eq.(\ref{t9}), we
have%
\begin{align}
W\left( \alpha ,\beta \right) & =A_{1}N_{m,n}^{-1}e^{2(\left\vert \alpha
\right\vert ^{2}+\left\vert \beta \right\vert ^{2})}\frac{\partial ^{2m+2n}}{%
\partial \tau ^{m}\partial t^{m}\partial \tau ^{\prime n}\partial t^{\prime
n}}\int \frac{d^{2}z_{1}d^{2}z_{2}}{\pi ^{4}}  \notag \\
& \times \exp \left\{ -\left( 2-A_{3}\right) \left\vert z_{1}\right\vert
^{2}-\left( 2-A_{3}\right) \left\vert z_{2}\right\vert ^{2}\right.  \notag \\
& +\left( t-2\alpha ^{\ast }+A_{2}z_{2}\right) z_{1}+\left( 2\alpha +\tau
+A_{2}z_{2}^{\ast }\right) z_{1}^{\ast }  \notag \\
& \left. \left. +2\left( \beta z_{2}^{\ast }-\beta ^{\ast }z_{2}\right)
+\tau ^{\prime }z_{2}^{\ast }+t^{\prime }z_{2}\right\} \right\vert
_{t,t^{\prime },\tau ,\tau ^{\prime }=0}  \notag \\
& =W_{0}\left( \alpha ,\beta \right) F_{m,n}\left( \alpha ,\beta \right) ,
\tag{C1}
\end{align}%
where $W_{0}\left( \alpha ,\beta \right) $ is defined in Eq.(\ref{t38}), and
\begin{align}
F_{m,n}\left( \alpha ,\beta \right) & =\frac{\left( -1\right) ^{m+n}}{N_{m,n}%
}\frac{\partial ^{2m+2n}}{\partial \tau ^{m}\partial t^{m}\partial \tau
^{\prime n}\partial t^{\prime n}}  \notag \\
& \times e^{R_{1}t+R_{2}\tau +R_{3}t^{\prime }+R_{4}\tau ^{\prime }}  \notag
\\
& \times \left. e^{K_{1}\left( \tau t+\tau ^{\prime }t^{\prime }\right)
+K_{3}\left( \tau \tau ^{\prime }+tt^{\prime }\right) }\right\vert
_{t,t^{\prime },\tau ,\tau ^{\prime }=0},  \tag{C2}
\end{align}%
and%
\begin{align}
R_{1}& =2\left( K_{1}\allowbreak \alpha -K_{3}\beta ^{\ast }\right)
=-R_{2}^{\ast },  \notag \\
R_{3}& =2\left( K_{1}\beta -K_{3}\alpha ^{\ast }\right) =-R_{4}^{\ast },
\tag{C3}
\end{align}%
as well as
\begin{equation}
K_{1}=\frac{\bar{n}+\cosh ^{2}r}{2\bar{n}+1},K_{3}=\frac{\sinh r\cosh r}{2n+1%
}.  \tag{C4}
\end{equation}%
Expanding the partial exponential items in Eq.(C2), then Eq.(C2) becomes

\begin{align}
& F_{m,n}\left( \alpha ,\beta \right)  \notag \\
& =\frac{\left( -1\right) ^{m+n}}{N_{m,n}}\frac{\partial ^{2m+2n}}{\partial
\tau ^{m}\partial t^{m}\partial \tau ^{\prime n}\partial t^{\prime n}}%
\sum_{l,j=0}^{\infty }\frac{K_{1}^{l+j}}{l!j!}  \notag \\
& \times \left. \left( \tau t\right) ^{l}\left( \tau ^{\prime }t^{\prime
}\right) ^{j}e^{R_{1}t+R_{3}t^{\prime }-R_{1}^{\ast }\tau -R_{3}^{\ast }\tau
^{\prime }+K_{3}\left( \tau \tau ^{\prime }+tt^{\prime }\right) }\right\vert
_{t,t^{\prime },\tau ,\tau ^{\prime }=0}  \notag \\
& =\frac{\left( -1\right) ^{m+n}}{N_{m,n}}\sum_{l,j=0}^{\infty }\frac{%
K_{1}^{l+j}}{l!j!}\frac{\partial ^{2l+2j}}{\partial \left( -R_{1}^{\ast
}\right) ^{l}\partial R_{1}^{l}\partial \left( -R_{3}^{\ast }\right)
^{j}\partial R_{3}^{j}}  \notag \\
& \left. \frac{\partial ^{2m+2n}}{\partial \tau ^{m}\partial t^{m}\partial
\tau ^{\prime n}\partial t^{\prime n}}e^{R_{1}t+R_{3}t^{\prime }-R_{1}^{\ast
}\tau -R_{3}^{\ast }\tau ^{\prime }+K_{3}\left( \tau \tau ^{\prime
}+tt^{\prime }\right) }\right\vert _{t,t^{\prime },\tau ,\tau ^{\prime }=0}.
\tag{C5}
\end{align}%
Further using the generating function of two-variable Hermite polynomials,%
\begin{align}
& \left. \frac{\partial ^{m}}{\partial \tau ^{m}}\frac{\partial ^{n}}{%
\partial \upsilon ^{n}}e^{-A\tau \upsilon +B\tau +C\upsilon }\right\vert
_{\tau =\upsilon =0}  \notag \\
& =(\sqrt{A})^{m+n}H_{m,n}\left( \frac{B}{\sqrt{A}},\frac{C}{\sqrt{A}}%
\right) ,  \tag{C6}
\end{align}%
Eq.(C5) can be put into the following form%
\begin{align}
F_{m,n}\left( \alpha ,\beta \right) & =\frac{K_{3}^{m+n}}{N_{m,n}}%
\sum_{l,j=0}^{\infty }\frac{K_{1}^{l+j}}{l!j!}\frac{\partial ^{2l+2j}}{%
\partial \left( -R_{1}^{\ast }\right) ^{l}\partial R_{1}^{l}\partial \left(
-R_{3}^{\ast }\right) ^{j}\partial R_{3}^{j}}  \notag \\
& \times H_{m,n}\left( \frac{R_{1}}{\sqrt{-K_{3}}},\frac{R_{3}}{\sqrt{-K_{3}}%
}\right)  \notag \\
& \times H_{m,n}\left( \frac{-R_{1}^{\ast }}{\sqrt{-K_{3}}},\frac{%
-R_{3}^{\ast }}{\sqrt{-K_{3}}}\right) .  \tag{C7}
\end{align}%
Using the relation
\begin{equation}
\frac{\partial ^{l+k}}{\partial x^{l}\partial y^{k}}H_{m,n}\left( x,y\right)
=\frac{m!n!}{\left( m-l\right) !\left( n-k\right) !}H_{m-l,n-k}\left(
x,y\right) ,  \tag{C8}
\end{equation}%
thus we can obtain Eq.(\ref{t39}).

\bigskip

\textbf{Appendix D: Derivation of Eq.(\ref{t49})}

Using the displacement operator $D_{a}\left( \alpha \right) =e^{\left\vert
\alpha \right\vert ^{2}/2}e^{-\alpha ^{\ast }a}e^{\alpha a^{\dagger }}$ and $%
D_{b}\left( \beta \right) =e^{\left\vert \beta \right\vert ^{2}/2}e^{-\beta
^{\ast }b}e^{\beta b^{\dagger }}$ as well as the normally ordering form of
PA-TMSTS (\ref{t17}), the CF of PA-TMSTS is given by%
\begin{align}
& \chi _{E}\left( \alpha ,\beta \right) \left. =\right. \mathtt{tr}\left[
D_{a}\left( \alpha \right) D_{b}\left( \beta \right) \rho ^{SA}\right]
\notag \\
& =\frac{A_{1}e^{(\left\vert \beta \right\vert ^{2}+\left\vert \alpha
\right\vert ^{2})/2}}{N_{m,n}}\frac{\partial ^{2m+2n}}{\partial \alpha
^{m}\partial \beta ^{n}\partial \left( -\alpha ^{\ast }\right) ^{m}\partial
\left( -\beta ^{\ast }\right) ^{n}}  \notag \\
& \times \mathtt{tr}\left[ \colon e^{\alpha a^{\dagger }+\beta b^{\dagger
}-\alpha ^{\ast }a-\beta ^{\ast }b+A_{2}\left( a^{\dagger }b^{\dagger
}+ab\right) -A_{3}\left( a^{\dagger }a+b^{\dagger }b\right) }\colon \right] .
\tag{D1}
\end{align}%
In a similar way to derive Eq.(\ref{t18}), using Eqs.(A1) and (\ref{t18}),
one can directly obtain%
\begin{align}
\chi _{E}\left( \alpha ,\beta \right) & =\frac{e^{(\left\vert \beta
\right\vert ^{2}+\left\vert \alpha \right\vert ^{2})/2}}{N_{m,n}}\frac{%
\partial ^{2m+2n}}{\partial \alpha ^{m}\partial \beta ^{n}\partial \left(
-\alpha ^{\ast }\right) ^{m}\partial \left( -\beta ^{\ast }\right) ^{n}}
\notag \\
& \times e^{B_{1}\left( \alpha \left( -\alpha ^{\ast }\right) +\beta \left(
-\beta ^{\ast }\right) \right) +B_{2}\left( \alpha \beta +\left( -\alpha
^{\ast }\right) \left( -\beta ^{\ast }\right) \right) }.  \tag{D2}
\end{align}%
Taking the following transformations%
\begin{align}
\alpha & \rightarrow \alpha +\tau ,-\alpha ^{\ast }\rightarrow t-\alpha
^{\ast },  \notag \\
\beta & \rightarrow \beta +\tau ^{\prime },-\beta ^{\ast }\rightarrow
t^{\prime }-\beta ^{\ast },  \tag{D3}
\end{align}%
which leads to%
\begin{align}
& e^{B_{1}\left( \alpha \left( -\alpha ^{\ast }\right) +\beta \left( -\beta
^{\ast }\right) \right) +B_{2}\left( \alpha \beta +\left( -\alpha ^{\ast
}\right) \left( -\beta ^{\ast }\right) \right) }  \notag \\
& \rightarrow \exp \left[ -B_{1}(\left\vert \alpha \right\vert
^{2}+\left\vert \beta \right\vert ^{2})+B_{2}\left( \alpha \beta +\alpha
^{\ast }\beta ^{\ast }\right) \right]  \notag
\end{align}
\begin{align}
& \times \exp \left[ B_{1}\left( \tau ^{\prime }t^{\prime }+t\tau \right)
+B_{2}\left( t^{\prime }t+\tau ^{\prime }\tau \right) \right]  \notag \\
& \times \exp \left[ t\left( \alpha B_{1}-\allowbreak B_{2}\beta ^{\ast
}\right) +\tau \left( \beta B_{2}-B_{1}\alpha ^{\ast }\right) \right]  \notag
\\
& \times \exp \left[ \tau ^{\prime }\left( \alpha B_{2}-B_{1}\beta ^{\ast
}\right) +t^{\prime }\left( \beta \allowbreak B_{1}-B_{2}\alpha ^{\ast
}\right) \right] ,  \tag{D4}
\end{align}%
thus Eq.(D2) becomes Eq.\textbf{(}\ref{t49}\textbf{)}.


\bigskip

\begin{thebibliography}{99}
\bibitem{1} D. Bouwmeester, A. Ekert and A. Zeilinger, \textit{The Physics
of Quantum Information} (Springer-Verlag, Berlin, 2000).

\bibitem{2} J. Eisert, S. Scheel, and M. B. Plenio, Phys. Rev. Lett. 89,
137903 (2002).

\bibitem{3} G. Giedke and J. I. Cirac, Phys. Rev. A. 66, 032316 (2002).

\bibitem{4} J. Fiurasek, Phys. Rev. Lett. 89, 137904 (2002).

\bibitem{5} T. Opatrn\'{y}, G. Kurizki, and D.-G. Welsch, Phys. Rev. A 61,
032302 (2000).

\bibitem{6} A. Zavatta, S. Viciani, and M. Bellini, Science 306, 660 (2004).

\bibitem{7} A. Zavatta, S. Viciani, and M. Bellini,\ Phys. Rev. A 72, 023820
(2005).

\bibitem{8} H. Nha and H. J. Carmichael,\ Phys. Rev. Lett. 93, 020401 (2004).

\bibitem{8a} J. Wenger, R. Tualle-Brouri, and P. Grangier, Phys. Rev. Lett.
\textbf{92}, 153601 (2004).

\bibitem{9} M. S. Kim,\ J. Phys. B 41, 133001 (2008).

\bibitem{10} L. Y. Hu and H. Y. Fan,\ J. Opt. Soc. Am. B 25, 1955 (2008).

\bibitem{11} V. Parigi, A. Zavatta, M. S. Kim, and M. Bellini, Science 317,
1890 (2007).

\bibitem{12} S. Olivares, M. G. A. Paris, and R. Bonifacio,\ Phys. Rev. A
67, 032314 (2003).

\bibitem{13} A. Kitagawa, M. Takeoka, M. Sasaki, and A. Chefles,\ Phys. Rev.
A 73, 042310 (2006).

\bibitem{14} A. Ourjoumtsev, A. Dantan, R. Tualle-Brouri, and P. Grangier,\
Phys. Rev. Lett. 98, 030502 (2007).

\bibitem{15} A. Ourjoumtsev, R. Tualle-Brouri, and P. Grangier,\ Phys. Rev.
Lett. 96, 213601 (2006).

\bibitem{16} L. Y. Hu, X. X. Xu and H. Y. Fan, J. Opt. Soc. Am. B \textbf{27}%
, 286 (2010).

\bibitem{17} P. T. Cochrane, T. C. Ralph, and G. J. Milburn,\ Phys. Rev. A
65, 062306 (2002).

\bibitem{18} S. D. Bartlett and B. C. Sanders,\ Phys. Rev. A 65, 042304
(2002).

\bibitem{19} M. Sasaki and S. Suzuki,\ Phys. Rev. A 73, 043807 (2006).

\bibitem{20} C. Invernizzi, S. Olivares, M. G. A. Paris, and K. Banaszek,\
Phys. Rev. A 72, 042105 (2005).

\bibitem{21} S. Y. Lee, S. W. Ji, H. J. Kim, and H. Nha, Phys. Rev. A 84,
012302 (2011).

\bibitem{22} L. Y. Hu, and Z. M. Zhang,\ J. Opt. Soc. Am. B 29, (2012) to be
published. or arXiv:1110.6587[quannt-ph]

\bibitem{23} G. S. Agarwal and K. Tara, Phys. Rev. A \textbf{43}, 492
(1991); Phys. Rev. A \textbf{46}, 485 (1992).

\bibitem{24} S. M. Barnett and P. M. Radmore, Methods in Theoretical Quantum
Optics (Clarendon Press, 1997).

\bibitem{25} M. O. Scully and M. S. Zubairy, Quantum Optics (Cambridge
University Press, 1998).

\bibitem{26} V. V. Dodonov,\ J. Opt. B 4, R1 (2002).

\bibitem{27} H Y Fan, H. L. Lu and Y. Fan, Ann. Phys. 321, 480 (2006).

\bibitem{28} FanHong-Yi, H. R. Zaidi, and J. R. Klauder,\ Phys. Rev. D 35,
1831 (1987).

\bibitem{29} R. R. Puri, Mathematical Methods of Quantum Optics
(Springer-Verlag, 2001), Appendix A.

\bibitem{30} H. Y. Fan, L.Y. Hu,\ Opt. Lett. 33, 443 (2008).

\bibitem{31} P. Marian, Phys. Rev. A 45, 2044 (1992).

\bibitem{32} P. Marian, T. A. Marian and H. Scutaru, J. Phys. A: Math. Gen.
34, 6969 (2001).

\bibitem{33} Z. X. Zhang and H. Y. Fan, Phys. Lett. A 174, 206 (1993).

\bibitem{34} W. M. Zhang, D. F. Feng, and R. Gilmore,\ Rev. Mod. Phys. 62,
867 (1990).

\bibitem{35} M. G. Benedict and A. Czirjak, Phys. Rev. A 60, 4034 (1999).

\bibitem{36} C. T. Lee, Phys. Rev. A 44, R2775 (1991).

\bibitem{37} J. K. Asboth, J. Calsamiglia, and H. Ritsch, Phys. Rev. Lett.
94, 173602 (2005).

\bibitem{38} E. Shchukin, W. Vogel,\ Phys. Rev. Lett. 95, 230502 (2005).

\bibitem{39} D. N. Klyshko,\ Phys. Lett. A 213, 7 (1996).

\bibitem{40} C. T. Lee,\ Phys. Rev. A 41, 1569 (1990).

\bibitem{42} E. P. Wigner,\ Phys. Rev. 40, 749 (1932).

\bibitem{43} D. E. Browne, J. Eisert, S. Scheel, and M. B. Plenio, Phys.
Rev.A 67, 062320 (2003).

\bibitem{44} R. Garc\'{\i}a-Patr\'{o}n, J. Fiur\'{a}\v{s}ek, N. J. Cerf, J.
Wenger, R. Tualle-Brouri, and P. Grangier,\ Phys. Rev. Lett. \textbf{93},
130409 (2004).

\bibitem{45} X. Y. Chen, Phys. Lett. A 372, 2976 (2008).

\bibitem{46} S. L Braunstein and P. van Loock, Rev. Mod. Phys. 77, 513
(2005).

\bibitem{47} P. Marian and T. A. Marian, Phys. Rev. A 74, 042306 (2006).

\bibitem{48} S. L. Braunstein and H. J. Kimble, Phys. Rev. Lett. 80, 869
(1998).

\bibitem{49} Y. Yang and F. L. Li,\ Phys. Rev. A 80, 022315 (2009).

\bibitem{50} L. Y. Hu, X. X. Xu, Z. S. Wang, and X. F. Xu, Phys. Rev. A
\textbf{82}, 043842 (2010).
\end{thebibliography}
\end{document}